\documentclass[sigconf, nonacm]{acmart}
\newcommand\vldbdoi{XX.XX/XXX.XX}
\newcommand\vldbpages{XXX-XXX}
\newcommand\vldbvolume{XX}
\newcommand\vldbissue{X}
\newcommand\vldbyear{2026}
\newcommand\vldbauthors{Dong and Wang}
\newcommand\vldbtitle{\shorttitle}
\newcommand\vldbavailabilityurl{}
\newcommand\vldbpagestyle{plain}

\usepackage{graphicx}
\usepackage{booktabs}
\usepackage{amsfonts}
\usepackage{xcolor}
\usepackage{microtype}
\usepackage{enumitem}
\usepackage[most]{tcolorbox}
\usepackage{listings}

\lstdefinestyle{compactpython}{
  language=Python,
  basicstyle=\ttfamily\footnotesize,
  keywordstyle=\color{blue!60!black},
  stringstyle=\color{green!45!black},
  commentstyle=\color{gray}\itshape,
  showstringspaces=false,
  breaklines=true,
  breakatwhitespace=true,
  columns=fullflexible,
  keepspaces=true,
  xleftmargin=1em,
  aboveskip=4pt,
  belowskip=4pt,
}
\newcommand{\shorttitle}{Compilation-Based Execution of Semantic Operators}

\title{From Interpretation to Compilation: Compilation-Based Execution of
Semantic Operators~[Vision]}

\author{Wenkai Dong}
\affiliation{%
  \institution{University of Hawaii at Manoa}
}
\email{dongw@hawaii.edu}

\author{Yifan Wang}
\affiliation{%
  \institution{University of Hawaii at Manoa}
}
\email{yifanw@hawaii.edu}

\begin{document}

\begin{abstract}
Semantic operator systems extend data processing with
natural-language interfaces, inventing semantic operations such as semantic filtering, mapping, and joining. In these systems, users can express predicates and transformations such as ``find beginner-friendly machine learning courses''. Existing implementations commonly execute such operators through an interpretation-based strategy: for each row, record, or candidate pair, the system invokes a large language model (LLM) to interpret the semantic intent and produce an output or judgement. While expressive, this execution model places expensive LLM calls inside the data-processing loop, leading to high latency, high monetary cost, and limited scalability.

In this vision paper, we propose a different execution paradigm:
compilation-based execution of semantic operators. Instead of using the LLM
as a runtime interpreter for every data item, we use the LLM once during the whole compilation phase to translate a semantic operator specification into deterministic executable code. The generated code acts as a compiled physical operator that \emph{mimics} the behavior of the original LLM-based semantic operator.
The system then applies this code locally over the dataset, avoiding per-row
or per-pair LLM calls during execution.
We instantiate this idea for semantic filter, semantic map, and semantic join operators. We compare compilation-based execution against LLM-based interpreted execution (baseline) and integrate the approach into an existing semantic operator system. Our preliminary results show that compilation-based execution can substantially reduce execution time and LLM calls while preserving much of the output quality of interpretation-based execution. More broadly, we argue that semantic operator systems should evolve from treating LLMs primarily as runtime executors toward treating them as semantic compilers that generate efficient executable plans. This shift opens a new research direction at the intersection of database query processing, program synthesis, and LLM-powered data systems.
\end{abstract}

\maketitle

\pagestyle{\vldbpagestyle}
\begingroup\small\noindent\raggedright\textbf{PVLDB Reference Format:}\\
\vldbauthors. \vldbtitle. PVLDB, \vldbvolume(\vldbissue): \vldbpages, \vldbyear.\\
\href{https://doi.org/\vldbdoi}{doi:\vldbdoi}
\endgroup
\begingroup
\renewcommand\thefootnote{}\footnote{\noindent
This work is licensed under the Creative Commons BY-NC-ND 4.0 International License. Visit \url{https://creativecommons.org/licenses/by-nc-nd/4.0/} to view a copy of this license. For any use beyond those covered by this license, obtain permission by emailing \href{mailto:info@vldb.org}{info@vldb.org}. Copyright is held by the owner/author(s). Publication rights licensed to the VLDB Endowment. \\
\raggedright Proceedings of the VLDB Endowment, Vol. \vldbvolume, No. \vldbissue\ %
ISSN 2150-8097. \\
\href{https://doi.org/\vldbdoi}{doi:\vldbdoi} \\
}\addtocounter{footnote}{-1}\endgroup

\ifdefempty{\vldbavailabilityurl}{}{
\vspace{.3cm}
\begingroup\small\noindent\raggedright\textbf{PVLDB Artifact Availability:}\\
The source code, data, and/or other artifacts have been made available at \url{\vldbavailabilityurl}.
\endgroup
}

\section{Introduction}
\label{sec:intro}
With the power of Large Language Model (LLM), \emph{semantic operators} are proposed, which are data-processing operators whose behaviors are specified by natural-language (NL) predicates rather than explicit symbolic conditions.
For example, a semantic filter can filter a course table
to find ``beginner-friendly courses about machine learning''. 
Commonly used semantic operators include semantic filter (\texttt{sem\_filter}), semantic join (\texttt{sem\_join}), semantic map (\texttt{sem\_map}), semantic aggregation (\texttt{sem\_agg}) and so on.  
With a series of builtin, first-class semantic operators, semantic operator systems~\cite{patel2025lotus, liu2025palimpzest, zhu2025nirvana, wang2025unify, russo2025abacus, dai2024uqequeryengineunstructured} have emerged that enables native semantic query processing for complex data analytics. For example, LOTUS introduces semantic operators as a declarative interface that extends relational processing with AI-based operations over structured and unstructured data~\cite{patel2025lotus}. Palimpzest proposes a
declarative system for optimizing AI-powered analytics over  unstructured data~\cite{liu2025palimpzest}. These systems show that semantic operators are a promising abstraction for LLM-powered data analytics.

However, the mainstream execution strategy for semantic operators is expensive. Many implementations execute semantic operators using what we call
\emph{interpretation-based execution}. In interpretation-based execution, the
LLM is invoked at runtime to interpret the semantic operator for each data item (cell or pair of cells).
For example, a semantic filter such as
\texttt{sem\_filter("beginner-friendly machine learning courses")} will be executed by repeatedly sending the predicate together with one row to the LLM
and asking whether that row satisfies the predicate, until all rows are checked. Similarly, a semantic map will send each row to the LLM to obtain a transformed value, while a semantic
join may send each candidate pair of records to the LLM to determine whether the pair satisfies the semantic relationship indicated by the predicate.

In other words, the LLM sits inside the inner loop of query execution. A filter
or map is executed conceptually as follows:

\begin{lstlisting}[style=compactpython]
for row in table:
    output = LLM(semantic_condition, row)
\end{lstlisting}

A semantic join is even more expensive because the LLM may be invoked for many
candidate pairs:

\begin{lstlisting}[style=compactpython]
for l, r in candidate_pairs:
    output = LLM(join_condition, l, r)
\end{lstlisting}

This execution model is general, but it is also costly. Each LLM invocation
incurs latency, monetary cost, and possible rate-limit overhead. 
For filters and
maps, the number of LLM calls grows linearly with the number of rows. For joins,
the number of calls grows with the number of candidate pairs, which could be
much larger. As a result, interpretation-based semantic execution can become
prohibitively slow and expensive on large datasets. Our initial evaluation proves this. 

Inspired by the idea of compiling SQL into code to speed up query execution~\cite{hyper}, we propose our vision  
for a complementary execution strategy: natural-language semantic operators
should not always be treated as prompts to be repeatedly interpreted by LLMs at
runtime. They can also be treated as high-level declarative programs that are compiled into
lower-level executable code for accelerating the execution.

\begin{figure*}[t]
  \centering
  \includegraphics[width=\textwidth]{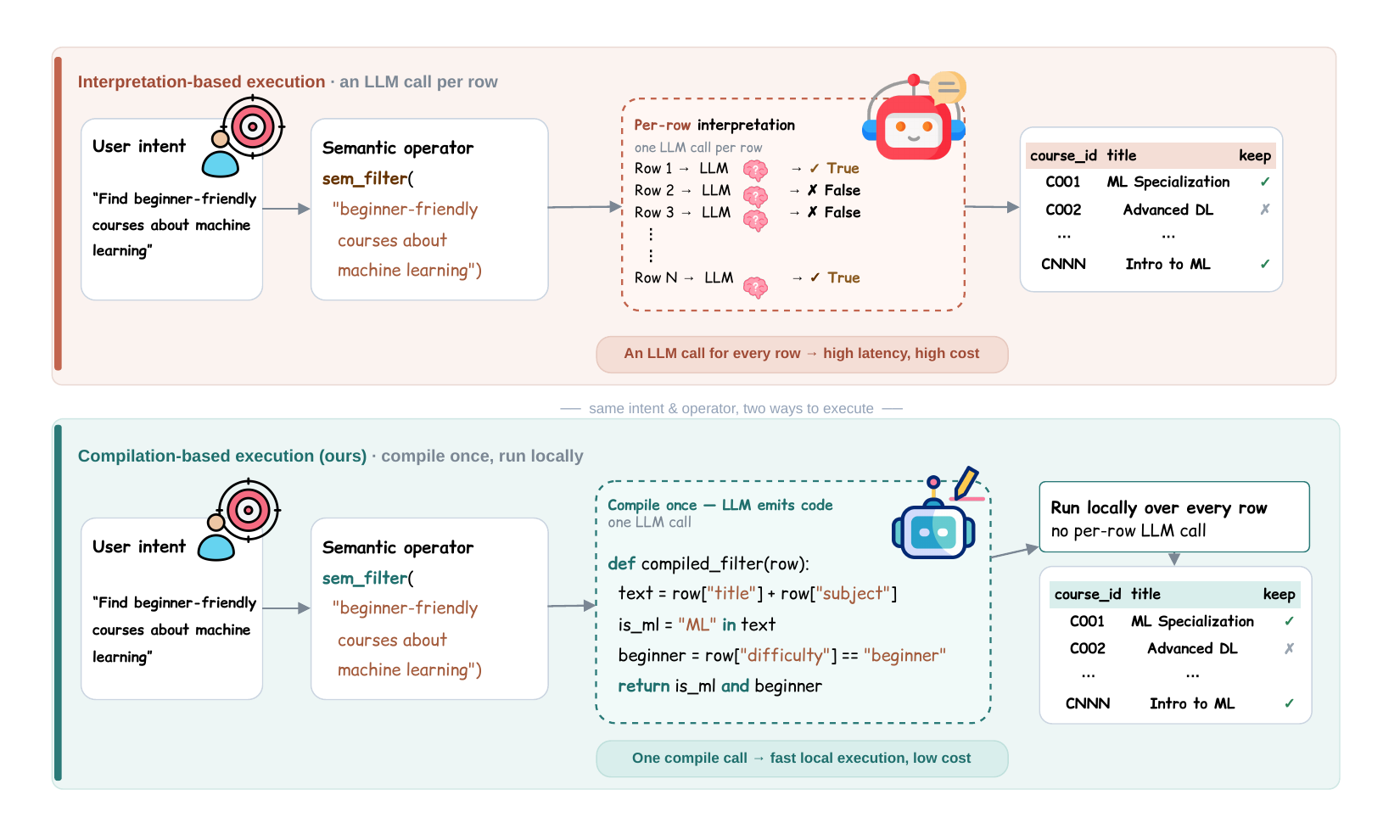}
  \caption{\textbf{Key idea:} move the LLM from per-row runtime interpretation to
  a one-time, compile-time synthesis of executable operator code. The interpreter
  (top) issues one LLM call per row; the compiler (bottom) issues a single LLM
  call to synthesize a deterministic function (\texttt{compiled\_filter} in the example), which then runs
  locally over every row with no per-row LLM call.}
  \label{fig:overview}
\end{figure*}

Specifically, we propose a new execution paradigm: \emph{compilation-based execution} for semantic operators. 
The key idea is to move the LLM out of the runtime loop and into an one-time compilation phase. Given a semantic operator and its predicate, the compilation-based system invokes the LLM once to
generate deterministic executable code that ``mimics'' the intended operator behavior. The generated code is then applied locally over the dataset without additional LLM calls.
Figure~\ref{fig:overview} illustrates this idea with a semantic filter example, {sem\_filter("beginner-friendly courses related to machine learning")}. 
An interpretation-based system (the upper workflow in the figure) asks the LLM to evaluate this predicate
separately for every row/course. In contrast, a compilation-based system (the lower workflow in figure) first asks
the LLM to synthesize a deterministic function that approximates the LLM evaluation logic for "beginner-friendly courses related to machine learning":

\begin{lstlisting}[style=compactpython]
def compiled_filter(row):
    text = (row["title"] + " " + row["subject"])
    is_ml = any(t in text for t in [
        "Machine Learning",
        "ML", 
        "Deep Learning",
        "DL", 
        "Artificial Intelligence",
        "AI"
    ])
    is_beginner = row["difficulty"] == "beginner"
    return is_ml and is_beginner
\end{lstlisting}

In the example, the generated function uses text fuzzy matching to approximate the LLM judging logic for whether the predicate is satisfied. The code in Figure~\ref{fig:overview} is a shorter version of the code above due to space limit. 
The generated function is then executed over all rows without  LLM calls. In this case, the LLM is used only
once to interpret the semantic intent and synthesize code, rather than
repeatedly to judge every row, avoiding the LLM call latency and cost per row. Therefore, compilation-based execution naturally takes less running time and less LLM cost than interpretation-based execution, especially on large datasets. The only problem is the approximation will unavoidably cause quality loss. Fortunately, our initial evaluation shows the loss is small.  
Table~\ref{tab:interpretation-vs-compilation} summarizes the difference between the two execution models.

\begin{table}[t]
\centering
\caption{Interpretation-based execution versus compilation-based execution of
semantic operators.}
\label{tab:interpretation-vs-compilation}
\small
\begin{tabular}{p{0.24\linewidth}p{0.33\linewidth}p{0.33\linewidth}}
\toprule
\textbf{Aspect} &
\textbf{Interpretation} &
\textbf{Compilation} \\
\midrule
Role of LLM &
Runtime executor &
Compiler \\

LLM calls &
Per row / candidate pair &
Once per operator \\

Execution target &
LLM API &
Local deterministic code \\

Strength &
Flexible reasoning &
Fast execution \\

Weakness &
High latency and cost &
Approximate behavior  \\
\bottomrule
\end{tabular}
\end{table}

This paper explores the feasibility and implications of compilation-based semantic operator execution. We instantiate the idea on three common semantic
operators: semantic filter, semantic map, and semantic join. For filters, the LLM
compiles a natural-language predicate into a Boolean function. For maps, it
compiles a natural-language transformation into a deterministic mapping
function. For joins, it compiles a semantic relationship into a pairwise function over two records. In all cases, the generated code is intended to
mimic the behavior of the corresponding interpretation-based LLM operator while
avoiding repeated LLM calls during execution.

Our current design is intentionally simple. The system performs a direct two-stage process: first
compile the semantic operator into code using an LLM, then execute the generated code locally. 
We evaluate this design in two settings. 
First, we compare compiled semantic
operators against naive LLM-based implementations that apply the LLM directly to
each row or candidate pair. Second, we integrate compilation-based execution
into an existing semantic operator system, Palimpzest~\cite{liu2025palimpzest} and compare end-to-end performance and
output quality with the original execution model in Palimpzest. Our results show that compilation-based execution can
significantly reduce runtime and LLM usage, while incurring only modest quality
degradation, proving compilation-based execution is not only a conceptual innovation but also a practical solution to improve semantic operators' speed and cost efficiency. Furthermore, we believe our vision and finding will inspire a broader range of research directions beyond semantic operators, making LLM-powered approaches more efficient and less costly in more areas.   

This paper makes the following contributions:

\begin{itemize}
    \item We propose a new semantic operator execution paradigm, \emph{compilation-based execution}, as a practical and effective solution to improve semantic operator system efficiency and cost usage.

    \item We implement a prototype of compilation-based execution and conduct initial evaluation, showing that it has significantly higher efficiency, lower cost, with only small approximation loss of quality. 

    \item We discuss the research challenges of semantic
    operator compilation, including operator compilability, semantic fidelity, and future optimizer integration. We further introduce how our vision may impact broader research areas. 
\end{itemize}

The broader message of this paper is that LLM-powered applications should not only optimize how LLM calls are scheduled, batched, cached, or pruned. They should also ask when LLM calls can be removed from execution entirely. 


\section{Related Work}
\label{sec:related}

\noindent\textbf{Semantic Operator Systems: } 
Recent work has proposed semantic operators as a declarative abstraction for
LLM-powered data processing. LOTUS introduces semantic operators that extend
relational-style processing with operations such as semantic filtering, joining,
aggregation, and ranking over structured and unstructured data
\cite{patel2025lotus}. 
Palimpzest similarly proposes
a declarative system for AI-powered analytics over arbitrary collections of
unstructured data, allowing users to specify high-level analytical workflows
while the system searches over model choices, prompts, and physical
implementations \cite{liu2025palimpzest}. Abacus further develops this direction
by introducing a cost-based optimizer for semantic operator systems, where
operators such as maps, filters, and joins may have multiple physical
implementations with different cost, latency, and quality tradeoffs
\cite{russo2025abacus}.
Our work is complementary to these systems. They
primarily optimize how LLM-based operators are executed, for example through
planning, model selection, prompt selection, batching, caching, etc. In contrast, we study whether the LLM can be moved out of the runtime execution loop entirely. 

\noindent\textbf{Database Query Compilation: }
Some modern query engines compile query plans into low-level code to reduce
interpretation overhead and improve CPU efficiency. For example, HyPer compiles
query plans into efficient machine code using LLVM, showing that compilation can substantially improve query execution performance on modern hardware
\cite{hyper}.
Our work is inspired by this database tradition, but applies the interpretation
versus compilation distinction to semantic operators. In traditional query
compilation, the input is a formal query with precise relational semantics, and
the compiler preserves exact query meaning. In our setting, the input is a
natural-language semantic operator whose behavior is typically defined by an LLM oracle or by the user's intended meaning. As a result, compilation is
approximate: the generated code is intended to mimic the behavior of the
interpretation-based LLM operator, but it may not be semantically equivalent.
This creates new research questions around semantic fidelity, operator compilability, and quality--cost tradeoffs.

\section{From Interpreter to Compiler}
\label{sec:model}

\paragraph{The interpretation-based execution model.}
A semantic operator is specified by a natural-language string, like a filter
predicate in semantic filter (\texttt{sem\_filter}), a per-row transformation in semantic map (\texttt{sem\_map}), or a join condition in semantic join (\texttt{sem\_join}). In today's
systems~\cite{liu2025palimpzest,patel2025lotus}, the operators are
\emph{interpreted}: at execution time the engine forwards each data item (row or pair),
together with the predicate, to an LLM and reads back the per-item result. A
unary operator over $N$ items therefore issues $N$ LLM calls, while a semantic join
over collections of size $M$ and $N$ issues up to $M\!\times\!N$. 
So both latency and money
cost grow with record cardinality and width. This execution model is expressive but slow,
expensive, and hard to scale, where the LLM calls are the bottleneck.

\paragraph{The key observation.}
For a large class of operators the LLM's per-item behavior is not open-ended, instead, many operators output a fixed-scope result based on some implicit attributes of the data item. For example, \texttt{sem\_filter} and \texttt{sem\_join} output binary judgment (True/False) on the current row or pair of rows; \texttt{sem\_filter("popular courses")} may implicitly relies on thresholding attribute ``course rating'' to determine the course popularity.  
Such semantic operators with close-ended output and relatively clear decision rules are possible to be simulated with a deterministic program, since they
are semantic on the surface but mechanical underneath.
For such operators, invoking an LLM once per item re-derives the same rule $N$ times in total.

\paragraph{Compile once, execute locally.}
Therefore, for such operators, we propose to use the LLM as a \emph{compiler} rather than an
interpreter. During a one-time compilation phase, a single LLM call
extracts the operator's implicit decision rules into a explicit, deterministic Python function, such that no future LLM calls are needed to re-derive the rules again and again. 
The function is
re-used for every item by local execution without LLM. Execution cost collapses from $O(N)$
(or $O(M\!\times\!N)$) LLM calls to a single compile call. The running time is therefore also reduced due to drastically less LLM calls. 

\section{Implementation}
\label{sec:operators}

\paragraph{Compiled operators.}
We instantiate compilation for the common per-item semantic operators, i.e., semantic filter, map and join, whose decision rules are easier to extract than multi-item operators like semantic aggregation.  
Compiled filter (\texttt{compiled\_filter(row)}) receives each record and outputs True/False on whether the record satisfied the extracted filter rule. Compiled map (\texttt{compiled\_map(row)}) transforms/maps each record to a new value based on the extracted mapping rule. Compiled join (\texttt{compiled\_join(row\_l, row\_r)}) receives a pair of records and decides True/False on whether the two records match based on the extracted join rule, and only matched records are kept as results.   
In every case a single compile prompt asks the LLM to emit a named deterministic function, 
and forbids any LLM call inside the generated function.
Figure~\ref{fig:compile-prompt} shows this compilation prompt for \texttt{sem\_filter} to \texttt{compiled\_filter} on the running example. The prompt is parameterized only by the
natural-language predicate and the row schema.

\begin{figure}[t]
\centering
\begin{tcolorbox}[enhanced, colback=black!3, colframe=black!55,
  boxrule=0.6pt, arc=3pt, left=6pt, right=6pt, top=4pt, bottom=5pt,
  fonttitle=\bfseries\small, coltitle=black, colbacktitle=black!10,
  title={Compile prompt~~\textnormal{(semantic \texttt{filter})}}]
\small
You are a database query compiler. Translate the semantic predicate
below into one deterministic, optimized Python function.

\medskip
\begin{tabular}{@{}p{0.20\linewidth}p{0.72\linewidth}@{}}
\textbf{Predicate} & \emph{``beginner-friendly courses related to machine
learning''} \\[3pt]
\textbf{Row schema} & \texttt{title, subject, difficulty} \\
\end{tabular}

\medskip
\textbf{Rules}
\begin{enumerate}[leftmargin=1.5em,itemsep=1pt,topsep=2pt,parsep=0pt]
\item Name it exactly \texttt{compiled\_filter(row)}.
\item Return a boolean (\texttt{True}/\texttt{False}).
\item Output \emph{only} valid Python---no markdown, no prose.
\item Do not call an LLM inside the generated code.
\end{enumerate}
\end{tcolorbox}
\caption{The compile prompt for a semantic \texttt{filter}, on the running
example. 
The same template (with different function names and return types) compiles the other operators, i.e., \texttt{compiled\_map} and
\texttt{compiled\_join}.}
\label{fig:compile-prompt}
\end{figure}

\paragraph{The compiler sees global information}
Compilation changes the information available to the LLM. 
The LLM interpreter can see just the current row or pair of rows, while LLM compiler can see more, including a few more samples of rows from the table. This is reasonable as compiler only call LLM once, so adding a few more samples to compilation prompt does not affect the overall cost and latency too much, and this can further improve the compiled function quality.  
As a test, when compiling \texttt{sem\_join}, we additionally include several real sample
rows from left and right tables in the compiler prompt, so that the model can observe
value-distribution overlap and better choose a correct join key. 

\section{Preliminary Results}
\label{sec:prelim}

We evaluate compilation-based execution in three settings of increasing
integration. First, we run each compiled operator as a \emph{standalone}
implementation, outside any engine, to isolate the per-operator behavior of
compilation versus per-item interpretation (\S\ref{sec:prelim:standalone}).
Second, we integrate the compiled operators into Palimpzest and run each
operator \emph{inside the engine},
comparing with the engine's native LLM-interpretation operator
(\S\ref{sec:prelim:inengine}). Third, instead of single operators, we run the pipelines consisting of multiple operators implemented for the complex
queries in \emph{SemBench}~\cite{lao2025sembench} (\S\ref{sec:prelim:sembench}), a benchmark specifically for semantic operator systems.

Unless otherwise noted, all settings use \texttt{gpt-5-mini} for both compilation method and interpretation baseline.
The first two single-operator evaluations have no pre-labeled ground truth, as there is no such dataset and we have to construct predicates for each operator ourselves, without labors to annotate the ground truth results. Therefore we treat the
results of the interpretation approach as the reference and report \emph{agreement}---the
fraction of items (rows or pairs of rows) on which the compiled operator returns the same result as the
interpreter. The third evaluation on SemBench (\S\ref{sec:prelim:sembench}) ships ground truth,
so we score our method and the baseline directly against the ground truth. 

\subsection{Standalone Operators (outside the engine)}
\label{sec:prelim:standalone}

We first implement and evaluate the three per-item operators, i.e., filter, map, and
join, outside any existing semantic operator systems. The compiled and interpreted versions are both run and the latter is used as ground truth. 
Table~\ref{tab:prelim-standalone} reports
agreement (accuracy, precision, recall, F1) between compiled operator results and those of the ground truth. 
The latency is also reported in three parts: the interpreted operator end-to-end latency ("Base"), the compilation phase latency ("Comp") and the execution latency of the compiled operator ("Exec"). 

\begin{table}[t]
\centering
\small
\setlength{\tabcolsep}{3.5pt}
\caption{Standalone per-operator results: agreement with interpreted operator
 and mean per-query time . Each operator is tested over 100 queries (\#Q). Different queries have the same single-operator but different predicates.
}

\label{tab:prelim-standalone}
\begin{tabular}{l r rrrr rrr}
\toprule
 & & \multicolumn{4}{c}{\textbf{Agreement}}
   & \multicolumn{3}{c}{\textbf{Time per query}} \\
\cmidrule(lr){3-6}\cmidrule(lr){7-9}
\textbf{Operator} & \textbf{\#Q}
& \textbf{Acc} & \textbf{P} & \textbf{R} & \textbf{F1}
& \textbf{Base(s)} & \textbf{Comp(s)} & \textbf{Exec(ms)} \\
\midrule
Filter & 100 & 0.914 & 0.975 & 0.887 & 0.890 & 130.0 & 28.3 & 1.34 \\
Map    & 100 & 0.912 & 0.911 & 0.919 & 0.911 & 138.7 & 15.5 & 2.43 \\
Join   & 100 & 0.994 & 0.970 & 0.980 & 0.961 & 703.4 & 22.6 & 4.23 \\
\bottomrule
\end{tabular}
\end{table}

The compiled operators agree with the interpreted implementations on the large majority of
data items: 91\% (filter), 91\% (map), and 99\% (join) accuracy and mostly more than 90\% F1. 
The efficiency improvement is significant: once
compiled, each query (with only one operator) is evaluated in microseconds, and the total time of compiled operator (compilation time + execution time) has a up to 31x speedup to interpreted version. 
These prove that compilation execution achieves large acceleration with tiny loss of quality in algorithm level without system level factors. 

\subsection{Evaluation of Single Operator inside Engine}
\label{sec:prelim:inengine}
We then integrates each compiled operator into Palimpzest and compare them with the engine's native LLM-backed operator
(\texttt{LLMFilter} / \texttt{LLMConvertBonded} / \texttt{NestedLoopsJoin})
on a suite of 30 / 36 / 29 filter / map / join queries over a movie dataset.
Here each query still include only a single operator, with unique predicate. Similar to standalone evaluation, we take the native operator's output as ground truth, and measure agreement, LLM-call count, dollar
cost, and end-to-end time. To make it more straightforward, Table~\ref{tab:prelim-inengine} reports agreement in the absolute values, while reports the LLM calls, cost, and time as ``how many times compilation saves in average'', i.e., the metric of native implementation over that of compiled operator.  

\begin{table}[t]
\centering
\small
\renewcommand{\arraystretch}{1.3}
\caption{Evaluation inside Palimpzest: agreement of the compiled operator
with the native LLM operator's output, and how many times the compilation saves in average on LLM calls, cost and time, by calculating native over compiled.}

\label{tab:prelim-inengine}
\begin{tabular*}{\columnwidth}{@{\extracolsep{\fill}} l r rr rrr}
\toprule
 & & \multicolumn{2}{c}{\textbf{Agreement}}
   & \multicolumn{3}{c}{\textbf{Native/Compiled}} \\
\cmidrule(lr){3-4}\cmidrule(lr){5-7}
\textbf{Op} & \textbf{\#Q} & \textbf{Acc} & \textbf{F1}
& \textbf{Calls} & \textbf{Cost} & \textbf{Time} \\
\midrule
Filter & 30 & 0.972 & 0.888 & 133x & 16.0x & 15.8x \\
Map    & 36 & 0.889 & 0.826 & 134x & 14.8x & 15.5x \\
Join   & 29 & 0.998 & 0.882 & 345x & 32.6x & 0.9x \\
\bottomrule
\end{tabular*}
\end{table}


The agreement of filter and join are still close to the ground truth, except map who has the lowest agreement (0.889 accuracy and 0.826 F1), caused by
the open-ended characteristic of \texttt{sem\_map}, which will be discussed later (\S\ref{sec:prelim:hard}).
The LLM usage and cost are significantly reduced (saving up to 31.6x cost). The time is largely saved except for \texttt{sem\_join}, where \texttt{compiled\_join} is slightly slower than native Palimpzest \texttt{sem\_join} (0.9x). This is because Palimpzest's has internal join parallelism, which improve its efficiency. However, this optimization does not save LLM cost as the number of calls remains unchanged. In contrast, compilation-execution could save both latency and cost, which is one of our biggest advantages.  

\subsection{Evaluation on SemBench Queries}
\label{sec:prelim:sembench}
To evaluate the performance on complex queries rather than single operator, we use the Palimpzest-based implementations of compiled and native operators to answer  
the ten movie queries (Q1--Q10) in SemBench. SemBench has provided ground truth operator pipeline to answer each query, with ground truth answer to each query. If the ground truth pipeline includes any operator that has no compiled version, we simply use a native operator. 
Since this benchmark has
ground truth answers, we score \emph{both} the baseline (full native operators) and our method (using compiled operators as much as possible) against the ground truth answers and report accuracy, F1, LLM calls, cost and time.

\begin{table*}[t]
\centering
\small
\renewcommand{\arraystretch}{1.15}
\caption{ SemBench movie pipelines Q1--Q10: all-native baseline vs.\
as-much-compiled (ours), scored against ground truth. ``\textemdash'' marks queries with no usable quality signal
(\S\ref{sec:prelim:sembench}). Under \emph{Time}, Native is the baseline's end-to-end time and Ours and Exec are ours' compilation and local-execution time. Last row: Acc and F1 are means over scored queries, other columns
totals.}

\label{tab:prelim-sembench}
\begin{tabular*}{\textwidth}{@{\extracolsep{\fill}} l rr rr rr rr rrr}
\toprule
& \multicolumn{2}{c}{\textbf{Acc}}
& \multicolumn{2}{c}{\textbf{F1}}
& \multicolumn{2}{c}{\textbf{LLM calls}}
& \multicolumn{2}{c}{\textbf{Cost (\$)}}
& \multicolumn{3}{c}{\textbf{Time}} \\
\cmidrule(lr){2-3}\cmidrule(lr){4-5}\cmidrule(lr){6-7}\cmidrule(lr){8-9}\cmidrule(lr){10-12}
\textbf{Q} & Native & Ours & Native & Ours & Native & Ours & Native & Ours
& Native\,(s) & Ours\,(s) & Exec\,(ms) \\
\midrule
Q1  & 1.000 & 1.000 & \textemdash & \textemdash & 125 & 1 & 0.033 & 0.006 & 262.4 & 24.1 & 12.1 \\
Q2  & 1.000 & 1.000 & 1.000 & 1.000 & 4 & 1 & 0.001 & 0.008 & 12.4 & 48.5 & 13.9 \\
Q3  & \textemdash & \textemdash & \textemdash & \textemdash & 4 & 1 & 0.000 & 0.005 & 25.2 & 24.6 & 21.0 \\
Q4  & 1.000 & 1.000 & 1.000 & 1.000 & 4 & 1 & 0.001 & 0.005 & 14.7 & 35.9 & 15.9 \\
Q5  & 0.500 & 0.900 & \textemdash & \textemdash & 100 & 1 & 0.035 & 0.007 & 9.3 & 23.3 & 26.9 \\
Q6  & 0.700 & 1.000 & \textemdash & \textemdash & 100 & 1 & 0.035 & 0.004 & 17.0 & 11.9 & 50.2 \\
Q7  & 0.647 & 1.000 & 0.667 & 1.000 & 100 & 1 & 0.034 & 0.005 & 17.0 & 15.0 & 34.1 \\
Q8  & \textemdash & \textemdash & \textemdash & \textemdash & 4 & 1 & 0.001 & 0.004 & 8.5 & 28.5 & 11.0 \\
Q9  & \textemdash & \textemdash & \textemdash & \textemdash & 10 & 1 & 0.000 & 0.007 & 72.8 & 46.6 & 0.0 \\
Q10 & \textemdash & \textemdash & \textemdash & \textemdash & 150 & 1 & 0.054 & 0.006 & 271.9 & 31.7 & 30.5 \\
\midrule
\textbf{Tot} & 0.808 & 0.983 & \textemdash & \textemdash & \textbf{601} & \textbf{10}
& \textbf{0.195} & \textbf{0.058} & \textbf{711.3} & \textbf{290.1} & \textbf{215.6} \\
\bottomrule
\end{tabular*}
\end{table*}

As reported in Table~\ref{tab:prelim-sembench}, the compiled operators are as accurate as
the baseline in most cases, and markedly \emph{more} accurate on the join-heavy queries:
on Q5--Q7 (which looks for pairs of data), compiled pipeline has significant higher quality than baseline (e.g., accuracy 0.90/1.00/1.00 VS 0.50/0.70/0.65), demonstrating that compilation execution not only effectively approximate the interpreted implementation, but also be able to improve the execution quality in some cases. The LLM calls and cost of compiled pipeline are always significantly lower than baseline, with a more than 2x speedup.  These further strengthen the effectiveness of our idea. 
The queries marked with `\textemdash'' are those cannot generate meaningful results, like both methods output no answers. So we exclude their quality scores. 


\subsection{Scaling with Input Size}
\label{sec:prelim:scaling}

The paradigm's key promise is that
compilation's benefit \emph{grows} with input size. To test this we use the Pzlimpzest-based implementation over different sizes of tables: filter and map vary the row
count ($N=30/50/100$), and the join varies the candidate-pair count ($N=100/400/900$). 


\begin{table}[t]
\centering
\small
\renewcommand{\arraystretch}{1.25}
\caption{Evaluation on varying table sizes in Palimpzest. $N$: number of rows or pairs; LLM Calls, Cost and End-to-End Time are reported as how many times is saved, i.e., native over compiled. F1 is again using native as ground truth. }
\label{tab:prelim-scaling}
\begin{tabular*}{\columnwidth}{@{\extracolsep{\fill}} l r rrr r}
\toprule
 & & \multicolumn{3}{c}{\textbf{Native/Compiled}} & \\
\cmidrule(lr){3-5}
\textbf{Op} & \textbf{$N$} & \textbf{Calls} & \textbf{Cost} & \textbf{Time} & \textbf{F1} \\
\midrule
Filter & 30  & 30x  & 4x  & 3.6x  & 0.89 \\
       & 50  & 50x  & 6x  & 6.5x  & 0.90 \\
       & 100 & 100x & 13x & 14.2x & 0.92 \\
\addlinespace
Map    & 30  & 30x  & 4x  & 4.0x  & 0.92 \\
       & 50  & 50x  & 7x  & 7.1x  & 0.91 \\
       & 100 & 100x & 14x & 12.9x & 0.91 \\
\addlinespace
Join   & 100 & 100x & 9x  & 0.4x  & 0.90 \\
       & 400 & 400x & 40x & 2.3x  & 0.91 \\
       & 900 & 900x & 91x & 5.6x & 0.85 \\
\bottomrule
\end{tabular*}
\end{table}

Table~\ref{tab:prelim-scaling} clearly confirms compilation execution's advantage grows with data scale. For any operator, when $N$ increases, average LLM cost saving increases rapidly (like from 8x to 90x saving on join), speedup becomes larger (like from 3.6x to 14.2x on filter), and quality remains high and stable (F1 keeps more than 0.9 in most cases). So large-scale datasets are especially suitable to use compilation execution. 

\subsection{Limitation of Compilation Execution}
\label{sec:prelim:hard}

Not every operator compiles cleanly. Operators whose semantics can reduce to
string, numeric, or key logic compile almost exactly, like join and filter. Compilation execution does not perform that well on operators which require open-ended world knowledge or judgment. In contrast, the predicates of map operator are usually more flexible and the output is relatively broader than join and filter, so map is often the lowest quality operator among the three implemented. But this does not affect the effectiveness of our idea as map still keep a high quality. 

Another limitation is the compilation phase unavoidably takes long time since the resulting code may be complex with many keywords and matching rules, so generating such long code becomes the bottleneck.  

\section{Research Opportunities}
\label{sec:agenda}

Compilation-based execution opens several questions at the intersection of query processing, program synthesis, and LLM systems, which we believe will inspire people from broader domains beyond semantic data systems. 

\begin{itemize}[leftmargin=1.2em,itemsep=2pt,topsep=2pt,parsep=0pt]
\item \textbf{Fidelity and verification.} When does a compiled operator faithfully reproduce its LLM counterpart, and can we estimate this \emph{before}
trusting it? Possible solutions include small-scale test on a validation set, attaching a confidence score to the generated code, and so on. 

\item \textbf{Compilability and operator routing.} Some predicates are
``implicitly deterministic'' and compile near-perfectly; others are irreducibly
semantic. The optimizer must decide, per operator, whether to compile, interpret, or both. Accurately estimating compilability is a big challenge and should be a first-class property alongside selectivity and cardinality in cost model.

\item \textbf{Hybrid and partial compilation.} Rather than one strategy per
operator, an engine could compile a predicate's deterministic skeleton and fall
back to the LLM only on the residual hard cases, i.e., letting lightweight compilation-based path handles more tasks while keeping the capability to use LLM only when necessary.

\item \textbf{Reuse, drift, and recompilation.} A compiled operator is a
reusable artifact, cacheable across queries and sessions, which raises classic
systems questions: when does data drift invalidate it, and how cheaply can it be
recompiled or repaired? These questions need more exploration. 

\item \textbf{Beyond tabular and text.} Multimodal operators (e.g., predicates
over images) are the hardest to compile to short deterministic code, as they are naturally more semantic than table and text. Whether
compilation extends to them is open. This probably requires synthesizing pipelines over learned models to extract and match visual features. 

\item \textbf{Compilation-based execution beyond semantic operators.} There are many other LLM-powered tasks satisfying (1) fixed-scope output and (2) decision rules is implicit but easy to extract, like spam filtering, document ranking, even LLM-based Text-to-SQL, which are good candidates for compilation execution. How to extend compilation-based execution to speed up those tasks beyond semantic data systems? This needs exploration. 
\end{itemize}

\section{Conclusion}
\label{sec:conclusion}
We argue that semantic operator systems should treat the LLM as a
\emph{semantic compiler} rather than a runtime interpreter: a single compilation call can translate a natural-language operator into deterministic code that runs locally, removing the LLM from the execution loop to reduce cost and time. 
Our initial results suggest that 
this preserves the interpreter's behavior while cutting LLM calls and cost by
one to two orders of magnitude and saving several times of latency. We finally introduce the research opportunities around query processing, program synthesis, LLM systems, and furthermore, the potential of our vision to impact broader areas.

\bibliographystyle{ACM-Reference-Format}
\bibliography{references}

\end{document}